\documentclass[aps,prl,twocolumn,superscriptaddress,amsmath,amssymb]{revtex4}
\usepackage{graphicx}
\usepackage{amsmath}
\newcommand{\be}{\begin{equation}}
\newcommand{\ee}{\end{equation}}
\newcommand{\bea}{\begin{eqnarray}}
\newcommand{\eea}{\end{eqnarray}}

\begin{document}

\title{Synthetic Gauge Field with Highly Magnetic Lanthanide Atoms}

\author{Xiaoling Cui}

\affiliation{Institute for Advanced Study, Tsinghua University,
Beijing, 100084,  P. R. China}

\author{Biao Lian}

\affiliation{Institute for Advanced Study, Tsinghua University,
Beijing, 100084,  P. R. China}
\affiliation{Department of Physics, Stanford University, Stanford, California 94305, USA }

\author{Tin-Lun Ho}
\affiliation{Institute for Advanced Study, Tsinghua University, Beijing, 100084,  P. R. China}
\affiliation{Department of Physics, The Ohio-State University, Columbus, Ohio, 43210, USA}

\author{Benjamin L. Lev}
\affiliation{Department of Physics, Stanford University, Stanford, California 94305, USA }
\affiliation{Department of Applied Physics, Stanford University, Stanford, California 94305, USA }
\affiliation{E. L. Ginzton Laboratory, Stanford University, Stanford, California 94305, USA }

\author{Hui Zhai}
\affiliation{Institute for Advanced Study, Tsinghua University, Beijing, 100084,  P. R. China}

\begin{abstract}

We present a scheme for generating a synthetic magnetic field and spin-orbit coupling via Raman coupling in highly magnetic lanthanide atoms such as dysprosium.  Employing these atoms offer several advantages for realizing strongly correlated states and exotic spinor phases.  The large spin and narrow optical transitions of these atoms allow the generation of synthetic magnetic fields an order of magnitude larger than those in the alkalis, but with considerable reduction of the heating rate for equal Raman coupling. The effective hamiltonian of these systems differs from that of the alkalis' by an additional nematic coupling term, which leads to a phase transition in the dressed states as detuning varies.  For \text{high-spin} condensates,  spin-orbit coupling leads to a spatially periodic structure,  which   is described in Majorana representation by a set of points moving periodically on a unit sphere. We name this a ``Majorana spinor helix" in analogy to the persistent spin-$\frac{1}{2}$ helix observed in electronic systems.

\end{abstract}

\maketitle

In the past few years, several groups have realized a synthetic magnetic field either in traps or in optical lattices~\cite{Ian_magnetic,Ian_Hall,Bloch_lattice,Sengstock_lattice}, and spin-orbit (SO) coupling ~\cite{Ian_SO, Ian_partialwave,Jing_SO,Shuai,Jing_fermion,MIT_fermion,Washtington} with alkali atoms.
These developments have highlighted  intriguing  physics in the ultracold atomic gas context.  
Vortices and the classical Hall effect have been observed with a Bose condensate exposed to a synthetic magnetic field~\cite{Ian_magnetic, Ian_Hall}.  SO-coupling in a Bose gas can lead to superfluid phases with stripe order~\cite{Zhai, Ho_Zhang} and a rich phase diagram~\cite{Ian_SO,Ho_Zhang,Stringari}, as well as  modify  the effective interaction between dressed-state atoms~\cite{Ian_partialwave}.   In addition, SO-coupling also leads to a divergent spin susceptibility, and the magnetic transition it implies has recently been observed~\cite{Shuai,Stringari2}. Recently, SO-coupled Fermi gases are found to display interesting spin dynamics, topological transitions of Fermi surfaces, and spin-dependent band structure~\cite{Jing_fermion, MIT_fermion}.

However, there are serious challenges with creating exotic quantum matter using the current  scheme of generating synthetic gauge fields. 
The small fine-structure splitting of the excited level used in the Raman-coupling process for alkalis leads to significant heating through spontaneous emission~\cite{Ian_PRA}. 
 While lowering the laser intensity will reduce heating, it will also  reduce the strength of the synthetic gauge field, pushing the
high-field regime of novel correlated physics beyond reach.  As of now, the number of vortices generated by synthetic magnetic field is far below that generated by rotations \cite{Ian_magnetic}.

We suggest the use of lanthanide atoms such as Dy~\cite{Dy,Dy_fermion} and Er~\cite{Er} to overcome these challenges.
The particular atomic structure of these atoms---narrow linewidth transitions, large ground-state orbital and spin angular momenta, and large fine-structure splitting---offer many advantages over the alkalis.   As we explain later, the narrow-line transitions (2-kHz wide in Dy versus 6 MHz in Rb) and the $L>0$ nature of the Dy ground and excited states provide a significant increase in Raman coupling without additional heating, or conversely, much less heating at a fixed Raman coupling.
In addition, the larger spin value of Dy ($F$=$J$=8 and $F$=21/2 for bosonic and fermionic isotopes, respectively) versus Rb ($F$=1) enhances the magnitude of the Berry's phase---hence the strength of the synthetic magnetic field---making the quantum Hall regime much more accessible.

Another important difference between lanthanides and alkalis lies within the effective single particle hamiltonian.  While alkalis' consist only of vector terms in spin space, lanthanides' possess sizable tensor terms that generate nematic order.
By varying the relative strength between vector and tensor terms---achievable via Raman and Zeeman detuning---one can induce a discontinuous transition between spinor states in the (dressed) ground state, an effect that can easily be observed.

For a spin-1 Bose gas,  SO-coupling can lead to a stripe phase which is a superposition of two spinor condensates with different momenta~\cite{Ian_SO,Ho_Zhang,Stringari}. For large-spin atoms such as Dy,  the ground state can be a superposition of several condensates with commensurate momenta, resulting in a spinor condensate periodic along the direction of SO-coupling.  The symmetries of these states  are conveniently described in Majorana representation as a set of  $2F$ points on a unit sphere~\cite{Majorana,non-abelian, Schwinger, highspins, highspins2, Biao}, each tracing a different trajectory loop over a period  in space --- a structure we refer to as a Majorana spinor helix in analogy to the persistent spin-$\frac{1}{2}$ helix observed in condensed matter systems \cite{ZhangHelix09}. Changes in SO-coupling can also induce transitions between Majorana helices of different symmetry.

\begin{figure}[t]
\includegraphics[width=3.5in]{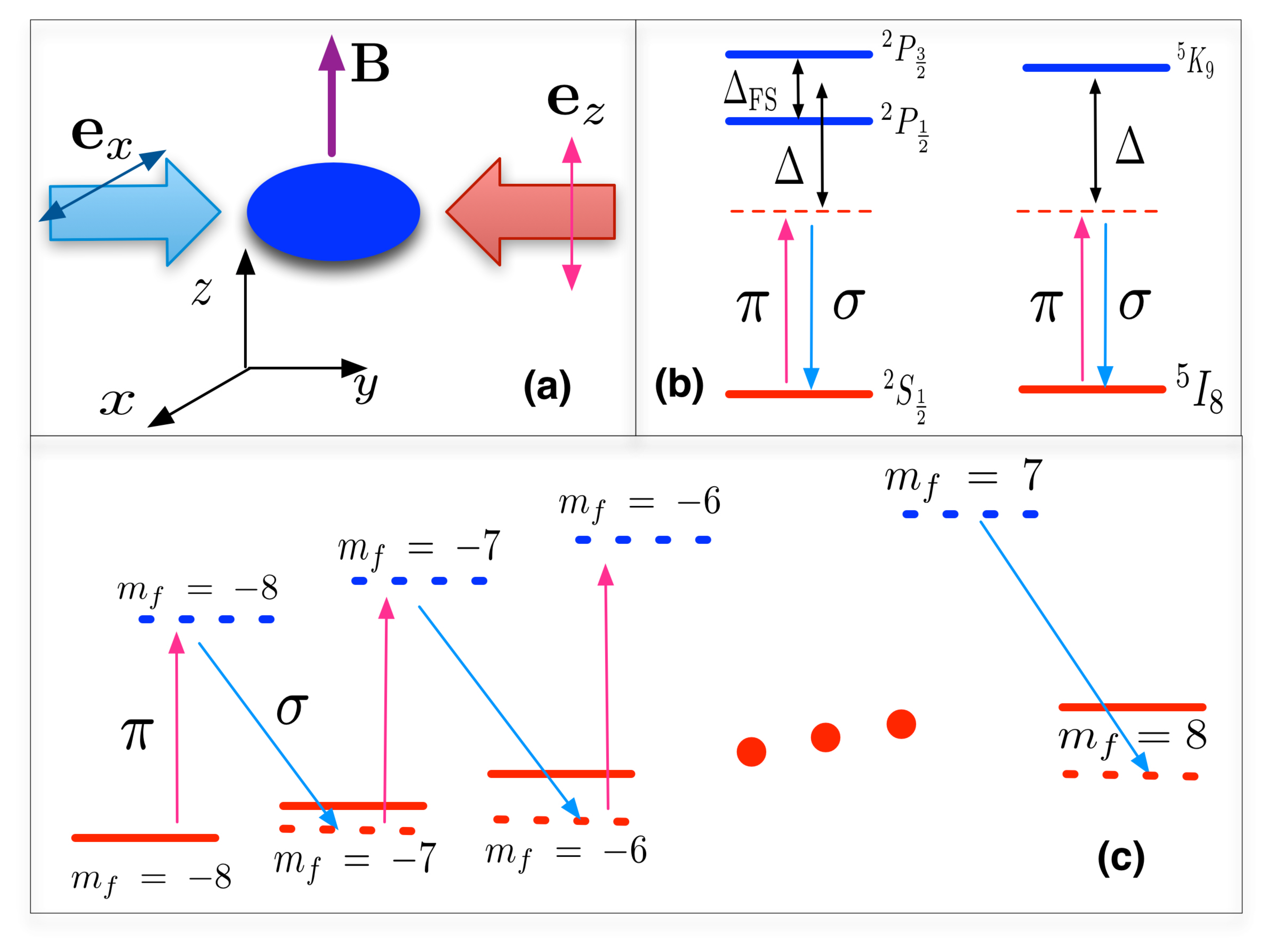}  
\caption{\label{Raman} (a) Raman laser and quantization B-field configuration; (b) Raman coupling scheme for an alkali atom like Rb (left) and for  Dy  (right); (c) Raman transition energy level diagram for magnetic sub-levels of the $F=8$ ground state of bosonic Dy.  \label{Dyfig} }
\end{figure}

{\it The effective hamiltonian of the Raman process: } 
As in Refs.~\cite{Ian_magnetic,Ian_SO,Ian_PRA}, we consider two-photon Raman coupling between different magnetic sub-levels of a spin-$f$ ground state. Coupling is via an excited state with spin-$f^\prime$ induced by an electric field ${\bf  {\cal E} }=\bf{E} + \bf{E}^{\dagger}$ in the presence of a magnetic field  $B$ along $\hat{\bf z}$,  as illustrated in Fig.~\ref{Dyfig}(a),  where 
 ${\bf  E}=g_{\sigma} \hat{\bf x} e^{i(k_{\sigma}y-\omega_{\sigma}t)} + g_{\pi} \hat{\bf z} e^{i(-k_{\pi} y-\omega_{\pi} t)}$.  
The first term  of ${\bf E}$ drives $\sigma_{+}$ and $\sigma_{-}$ transitions with respect to the $\hat{\bf z}$ spin quantization axis, while the second term drives the $\pi$ transition.  

To derive the effective hamiltonian for the $f$-spins under the Raman coupling, we extend the calculation in Ref.~\cite{Geremia} for  purely $\sigma^{\pm}$ light  to the configuration at hand. Eliminating the excited states, we obtain (in the same notation as Ref.~\cite{Geremia}) 
$\hat{H}_\text{R}={\bf E}^{\dagger} \sum_{f^\prime}\hat{\alpha}_{ff^\prime}/\Delta_{ff^\prime} {\bf E}^{}$, where $\hat{\alpha}_{ff'}= \hat{P}_f {\bf d} \hat{P}_{f'} {\bf d}^{\dag} \hat{P}_f $, ${\bf d}$ is the dipole operator, the $\hat{P}$'s are projectors onto  spin $f$ and $f'$, $\Delta_{ff'}=\omega_{\pi,\sigma}-\omega_{ff^\prime}$ is the detuning from the energy difference $\omega_{ff^\prime}$ between the $f$ and $f^\prime$ levels in a $\pi$ or $\sigma$ transition, and $\omega_{\pi, \sigma}$ is either $\omega_{\pi}$ or $\omega_{\sigma}$ depending on the transition.   We decompose $\hat{H}$ into scalar, vector, and tensor components (denoted as $\hat{H}_{\text{R}(0)}$, $\hat{H}_{\text{R}(1)}$, and $\hat{H}_{\text{R}(2)}$ respectively), 
\bea
\hat{H}_{\text{R}(0)}&=&\sum_{f'} \frac{1}{3}
\frac{\alpha_{ff'}^{(0)}g^2}{\Delta_{ff'}} \hat{I}_f \Big(
\hat{a}_{0}^{\dag}\hat{a}_0 +\hat{a}_+^{\dag}\hat{a}_+
+\hat{a}_-^{\dag}\hat{a}_-\Big), \nonumber \\
\hat{H}_{\text{R}(1)}&=&\sum_{f^\prime} \frac{1}{2} \frac{\alpha_{ff'}^{(1)}g^2}{\Delta_{ff'}} \Big\{
   \hat{F}_+ (-\hat{a}_0^{\dag}\hat{a}_+ A + \hat{a}_-^{\dag}\hat{a}_0 A^*) +\text{h.c.} \Big\}, \nonumber \\
H_{\text{R}(2)}&=&-\sum_{f'} \frac{\alpha_{ff'}^{(2)}g^2}{\Delta_{ff'}} \Big\{\hat{F}_+ (\hat{F}_z+\frac{\hat{I}_{f}}{2}) (\hat{a}_0^{\dag}\hat{a}_+ A + \hat{a}_-^{\dag}\hat{a}_0 A^*)  \nonumber\\
 & &+ \hat{F}_+^2  \hat{a}_-^{\dag}\hat{a}_+ +\text{h.c.} \Big\}, \label{H0}
\eea
where $A=e^{i[(k_{\sigma}+k_{\pi})x+\Delta\omega_\text{L} t]}$,  $a^{\dagger}_{\pm, 0}$ are photon creation operators for modes $\mp (\hat{\bf x}\pm i \hat{\bf y})$ and 
$\hat{\bf z}$ respectively, 
$\hat{F}_{\pm}=\mp(\hat{F}_x\pm i \hat{F}_y)/\sqrt{2}$,   $\hat{F}_{x,y,z}$ are spin-$f$ operators, $\hat{I}_f$ is the identity matrix, and $\Delta \omega_{L} = \omega_{\pi} - \omega_{\sigma}$. We have taken  $g_{\pi}=g_{\sigma}=g$ for simplicity.  $\alpha_{ff'}^{(i)} (i=0,1,2)$ are polarization constants 
given in Ref.~\cite{Geremia} corresponding to a rank-$i$ ($i=0,1,2$) Raman coupling in the ground state spin-$f$ manifold through an intermediate excited spin-$f^\prime$ state.

In deriving Eq.~(\ref{H0}), we have used two conditions: (i) the detuning $\Delta_{f,f'}$, is much larger than the difference in Zeeman energy both among spin-$f$ ground states and among spin-$f^\prime$ excited states, and thus one can ignore the Zeeman-shift-induced $m$ dependence in the detuning $\Delta_{ff'}$ \cite{deg};
and (ii) $\Delta \omega_{L} \ll \omega_\sigma, \omega_\pi$, and thus, one can set $\omega_{\pi, \sigma}=\overline{\omega}$, the mean value. Both conditions (i) and (ii) can be easily satisfied.

In addition, we assume (iii) that $\Delta \omega_{L}$ is close to the Zeeman energy $\omega_{z}$,  i.e., $\Delta \omega_{L} \sim \omega_{z}$, or $\delta \equiv \Delta \omega_L- \omega_{z} \ll \omega_z$. This condition, which is easily satisfied,  allows us to simplify $\hat{H}_{\text{R}(2)}$ by ignoring the $F_{+}^{2}$ term: the rotating wave approximation transforms away the time dependence of $A$ by performing a rotation $e^{-i\Delta \omega_{L} F_{z}t}$.  The $F^{2}_{+}$ term  gains a phase factor $e^{-2i\Delta \omega_{L}t}$, which  then averages to zero.

{\em Comparison between alkalis and open-shell lanthanides:}  For alkalis, the Raman transitions operate on the ground state $^{2}S_{1/2}$ and excited states 
 $^{2}P_{3/2}$ and $^{2}P_{1/2}$ possessing a fine-structure splitting $\Delta_{\text{FS}}$.  When the detuning is much larger than fine-structure splitting  
 $\Delta_{ff'}\gg  \Delta_{\text{FS}}$, 
$\hat{H}_{\text{R}(1)}$ is of order $g^2\Delta_{\text{FS}}/\Delta^{2}_{ff^\prime}$~\cite{Geremia,Deutsch}, where  $\Delta_{ff'} = \overline{\omega} -E_{\text{PS}}$ and $E_{\text{PS}}$ is the energy difference between the $P$ state  and the $S$ state without fine-structure splitting.  The reason is that the states $^{2}P_{3/2}$ and $^{2}P_{1/2}$ couple to the ground state $^{2}S_{1/2}$ through $\alpha^{(1)}$  with Clebsch-Gordon coefficients of opposite sign, and the $f'$-sum in $\hat{H}_{\text{R}(1)}$ in Eq.~(\ref{H0}) is then of the form  $\hat{H}^{\text{alkali}}_{\text{R}(1)}\sim g^2\left(\Delta_{ff'}^{-1} - (\Delta_{ff'} + \Delta_{\text{FS}})^{-1}\right) \sim g^2\Delta_{\text{FS}}/\Delta_{ff'}^2$.  
In contrast, in lanthanides such as Dy, the fine-structure splitting of excited state is very large. For instance, the 741-nm Dy Raman transition couples to a single excited state $^{5}$K$_{9}$ \cite{LevSpect}, resulting in a more favorable scaling $\hat{H}^{\text{Dy}}_{\text{R}(1)}\sim g^2/\Delta_{ff'}$.


Moreover, $\hat{H}_{\text{R}(2)}$  will vanish for the alkalis once $\Delta_{ff'}$ exceeds the hyperfine splitting. This is because the ground state is $J=1/2$, which has no matrix element through the 
rank-2 operator $\alpha^{(2)}$ back to the same manifold.  The hyperfine interaction, however, turns the ground state of $^{87}$Rb bosons into a spin $F=1$ particle, leading to  $\hat{H}^{\text{alkali}}_{\text{R}(2)}\sim g^2\Delta_{\text{HF}}/\Delta_{ff'}^2\ll \hat{H}^{\text{alkali}}_{\text{R}(1)}$.  In contrast, the ground state of Dy bosons and fermions have large orbital and spin angular momentum, and the matrix element $\alpha^{(2)}$ is non-zero within the ground state manifold, even without a hyperfine interaction: $\hat{H}^{\text{Dy}}_{\text{R}(2)}\propto g^2/\Delta_{ff'} \propto \hat{H}^{\text{Dy}}_{\text{R}(1)}$. 

The heating rates for both alkali and lanthanides are given by $ \Gamma \sim g^2 \gamma^i /\Delta_{ff'}^2$, where $\gamma^i$ is the excited state linewidth of atomic species $i$.  Since the linewidth of the $^{5}$K$_{9}$ excited state of Dy at 741~nm is $\sim$$10^{-3}$$\times$ narrower than that of alkali's relevant excited states, there is much less heating for Dy for the same amount of detuning and laser intensity.  On the other hand, 
${\Gamma}^{\text{Dy}}\sim (\gamma^{\text{Dy}}/\Delta_{ff'})\hat{H}_{\text{R}(1)}^{\text{Dy}} $,  and
$ {\Gamma}^{\text{alkali}} \sim ( \gamma^{\text{alkali}} / \Delta_{\text{FS}}^{\text{alkali}}) \hat{H}_{\text{R}(1)}^{\text{alkali}}$. Thus, for the same strength of Raman coupling, the ratio of two heating rates
$({\Gamma}^{\text{Dy}}/{\Gamma}^{\text{alkali}}) \sim ( \Delta_{\text{FS}}^{\text{alkali}} /\Delta_{ff'})(\gamma^{\text{Dy}} / \gamma^{\text{alkali}}) $ can be several orders of magnitudes smaller than unity for practicable laser intensities.

\begin{figure}[btp]
\includegraphics[width=3.5in]{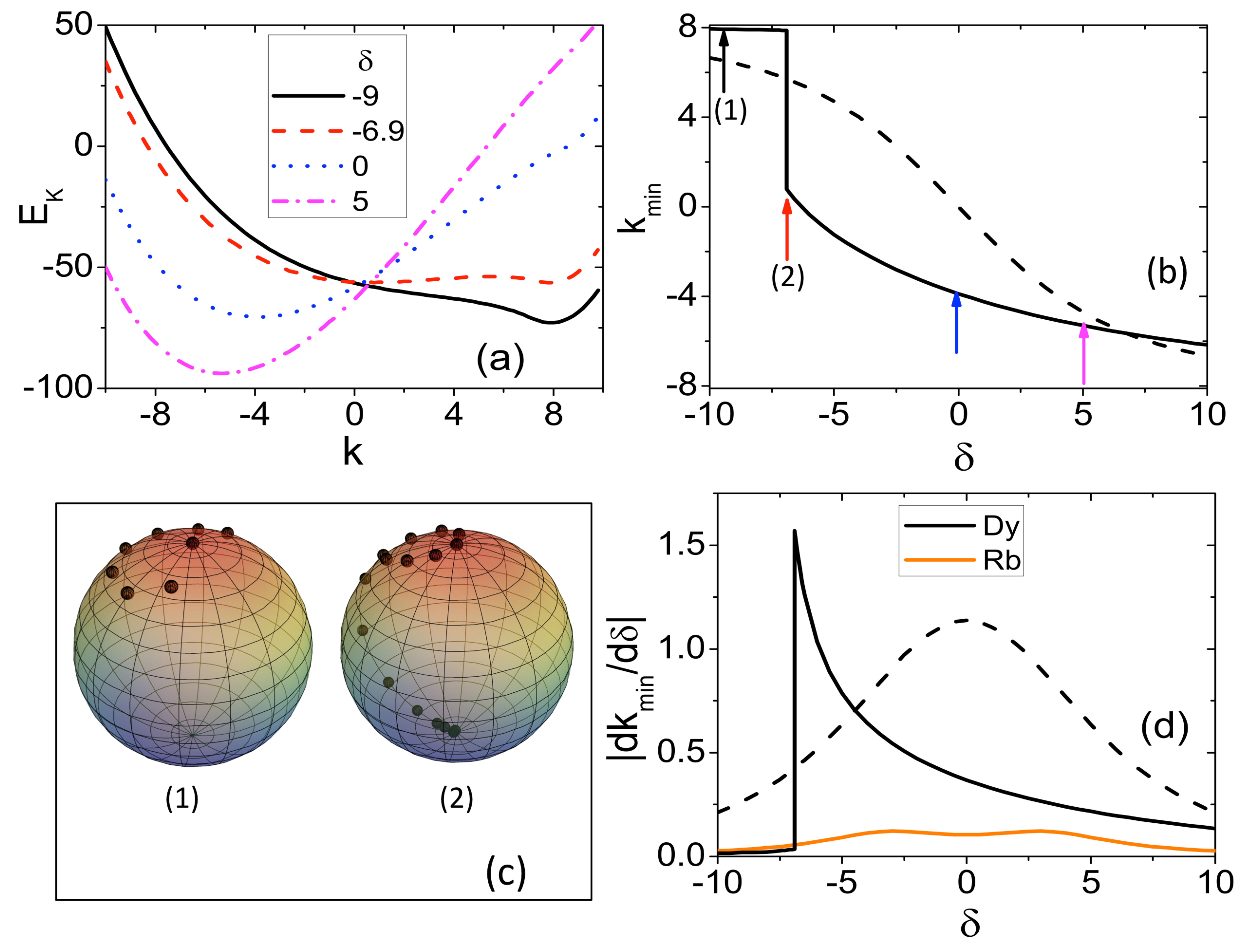}  
\caption{(a) The dispersion of the lowest branch for the Dy case with different $\delta$ [indicated by arrows in (b)]. (b) $k_{\text{min}}$ as a function of $\delta$ for Dy. The dashed line in (b) excludes the tensor term. (c) Majorana spinor representation for two ground-state spinor wavefunctions at different $\delta$ [marked by (1) and (2)] in (b). (d) $dk_{\text{min}}/d\delta$ (which is proportional to $B_{\text{eff}}$) for Dy (black solid line; black dashed line excludes tensor term contribution) and for Rb (orange solid line). In all cases, $\Omega$ is fixed at $4E_{\text{L}}$, and for Rb case the strength of quadratic Zeeman term $\omega_q F_z^2$ is taken as $\omega_q=1.9E_{\text{L}}$ as used in Ref. \cite{Ian_magnetic}. In all plots, [$k_{\text{min}}$, $E_{\text{k}}$, $\delta$] are in units of [$2k_\text{L}$, $E_{\text{L}}$, $E_{\text{L}}$].   \label{magnetic} }
\end{figure}

{\it The single particle hamiltonian for bosonic Dy.}
 With the photon fields replaced by their mean values $\langle a_{\pm, 0}\rangle$,  
Eq.~(\ref{H0}) can be simplified to
\begin{eqnarray}
H_{\text{R}}&=&  \Omega (e^{-i2k_L y}\hat{\Lambda}+ \text{h.c.}) + \delta F_{z}, \label{HR}\\
\hat{\Lambda}&=& \hat{F}_+ \big[ \hat{I}_{f}- 2C(\hat{F}_z+\hat{I}_{f}/2) \big], \label{Dy}
\end{eqnarray}
where $C$ is a constant that depends on $f'$ and $f$. For the 741-nm transition to the ${}^5K_9$ state of a ${}^{164}$Dy atom, we have $f^\prime=f+1$.   Using  the expressions for $\alpha^{(i)}_{f,f+1}$ ($i=1,2$)  given in Ref.~\cite{Geremia} and replacing $\hat{a}_{\pm,0}$ with their expectation value, we find  $\Omega=\frac{g^2\langle a_-^{\dag}\rangle \langle a_0\rangle}{2\Delta_{ff^\prime}\beta}  |\langle f ||{\bf r}|| f' \rangle|^2 $, 
where $\beta= (f+1)$ and  $C= 1/(2f+3)$.

The term $\delta F_z$ arises from the detuning of the frequency difference $\Delta \omega_{L}$ with the Zeeman frequency $\omega_z$.  Note, the tensor quadratic term $\hat{F}_{+}\hat{F}_{z}$ is absent in previous studies of alkali atoms \cite{Cr}. To understand the physics of  $\hat{H}_{R}$, it is useful to apply a spin rotation along $z$ to remove the phase factor in Eq.~(\ref{HR})~\cite{Ian_SO,Ho_Zhang}. The single particle hamiltonian along $\hat{y}$ is then
\begin{align}
&\hat{H}=\left(\frac{(\hat{k}_y-2k_\text{L}F_z)^2}{2M} + \delta F_z\right) + \Omega(\Lambda+\Lambda^{\dag}) \label{HDy}. 
\end{align}

{\it Abelian regime and synthetic magnetic field.} The abelian  regime corresponds to large $\Omega$ and $\delta$, and a synthetic magnetic emerges when 
there is field gradient $\delta = Gx$~\cite{Ian_magnetic,Ian_PRA}. In this regime,  Eq.~(\ref{HR}) at each point  ${\bf r}=(x,y,z)$ has a unique minimum $|\psi ({\bf r})\rangle$ in spin space separated from other excited states by an energy gap of order  $\Omega$.   Projecting $\hat{H}$ onto this lowest state, one obtains the hamiltonian resembling that of a charged particle in a magnetic field, where the magnetic field is given by ${\nabla}\times {\bf  A}$, with  the Berry's phase connection 
${\bf A} = -i\langle \Psi({\bf r})|\nabla|\Psi({\bf r})\rangle$ \cite{scalar}. A quick way to obtain the strength of the synthetic magnetic field near ${\bf r}=0$ is to calculate the spectrum of Eq.~(\ref{HDy}) in semi-classical limit, i.e., by calculating the spectrum for a given $\delta$ and expanding about its minimum $(0, k_{\text{min}}(\delta),0)$ along $k_y$: 
\begin{equation}
\mathcal{E}(k_y, \delta)=  \frac{1}{2M} (k_y-k_\text{min}(\delta))^2. 
\end{equation} 
The minimum $k_{\text{min}}$ can be regarded as $A_y$. The corresponding synthetic magnetic field is $B_{\text{eff}}=\partial A_y/\partial x=(\partial A_y/\partial \delta) (\partial\delta/\partial x) \propto dk_{\text{min}}/d\delta$ \cite{Ian_magnetic,Ian_PRA}. 

It is important to note that in the abelian limit, the ground state of $\hat{H}$ is controlled by two parameters $\Omega$ and $\delta$, which drive  the 
system towards the minima of $\hat{\Lambda}$ and $\delta \hat{F}_{z}$, respectively. As $\delta$ increases, 
 the ground state will change from one minimum to another. This can be seen in the minimum $k_{\text{min}}(\delta)$ of the spectrum $\mathcal{E}(k_y, \delta)$, which undergoes a jump at $\delta_\text{c}\cong -6.9$ as shown in  Fig.~\ref{magnetic}(a) and (b). 
 The nature of the ground state is shown in Fig.~\ref{magnetic}(c)  in Majorana representation. 
For  $\delta < \delta_{\text{c}}$, the system is ferromagnetic like, with all $2F=16$ Majorana points close to the north pole [see Fig.~2(c1)].  (A complete collapse onto the north pole results in a full ferromagnet.) As $\delta$ exceeds $\delta_{\text{c}}$, the state gains more nematicity, as seen from the migration of Majorana points towards the south pole [see Fig.~2(c2)]. (Maximum nematicity corresponds to equal distribution of Majorana points at both poles.)  The dashed line in Fig.~\ref{magnetic}(b) shows the behavior of 
$k_{\text{min}}$ without the tensor term. In this case, there is no discreet jump in $k_{\text{min}}$, which is similar to the case of $^{87}$Rb \cite{Ian_PRA}.

In Fig.~\ref{magnetic}(d) we compare $dk_{\text{min}}/d\delta$ (i.e., the $B_{\text{eff}}$ achieved for the same Zeeman field gradient)  between Dy and Rb. With the tensor term excluded (dashed line), the maximum value of $dk_{\text{min}}/d\delta$ achieved for Dy  is larger than that for Rb case by a factor of $\sim$$F=8$. 
This is because the larger spin structure enhances the maximum momentum transfer from $4 k_{L}$ in Rb to $4F k_{\text{L}}$ in Dy. 
Including the tensor term (black solid line), the synthetic magnetic field $dk_{\text{min}}/d\delta$ is greatly suppressed when $\delta<\delta_{\text{c}}$, but is much enhanced  for $\delta$ just above $\delta_{\text{c}}$.  Operating in the latter range of $\delta$ will provide a very strong synthetic magnetic field.

\begin{figure}[tp]
\includegraphics[width=3.4in]{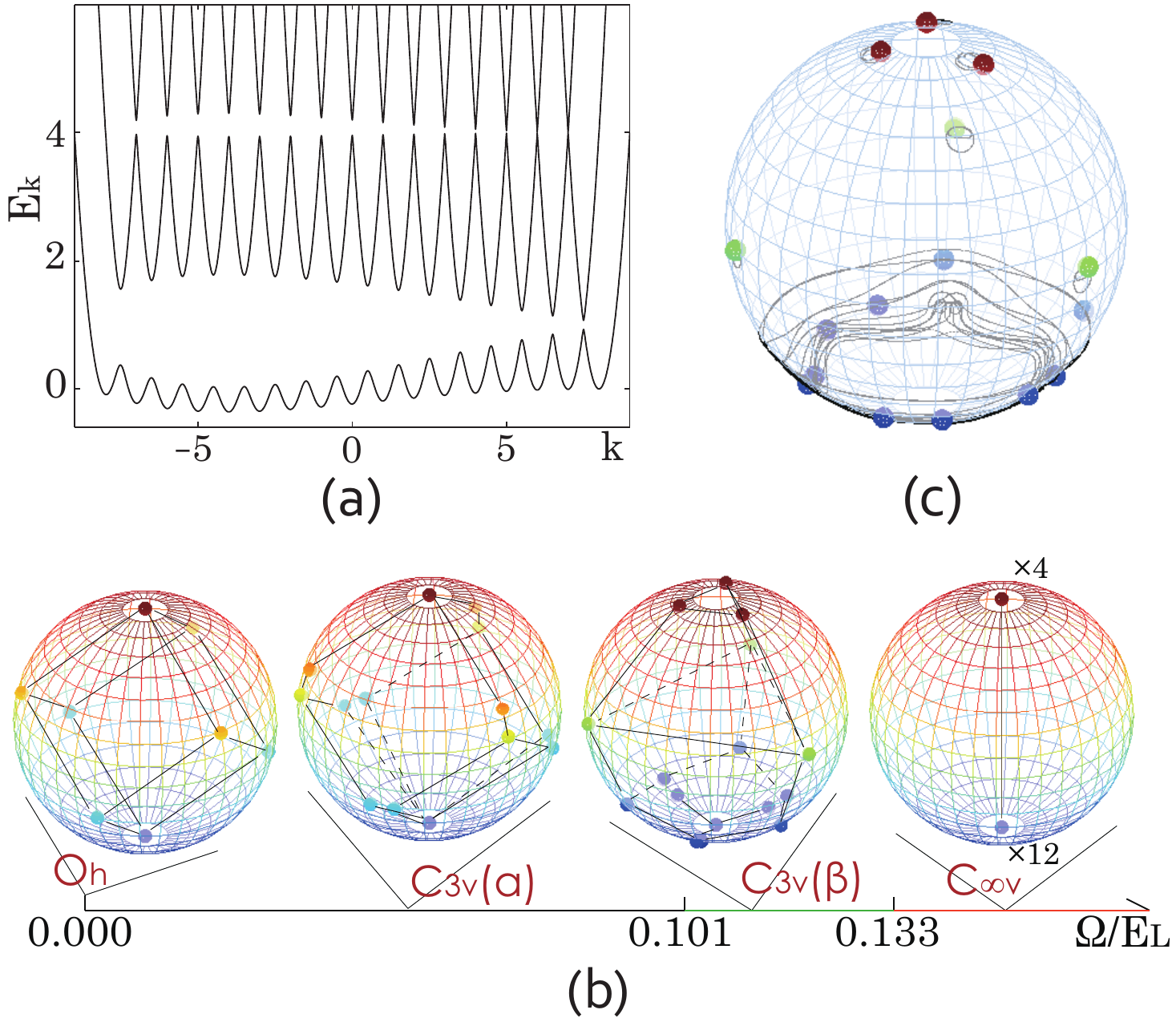}  
\caption{(a) Single particle spectrum for $\delta=0$ and $\Omega=0.083E_{\text{L}}$. $k$, $E_{k}$ are in  units of $2k_L$ and $E_L$, respectively. (b) A one-parameter phase diagram in term of $\Omega/E_{\text{L}}$. $\delta$ is fixed at zero. For the purpose of illustration, we choose a set of parameters $\{g_0,\dots,g_{16}\}=\{6,7,7,7,7,2,0,7,7\}$ as an example (other interaction parameters will result in spinors of differing symmetry, see Ref.~\cite{Biao}; these parameters remain unmeasured for Dy). With this set of interaction parameters, the system is in an octahedron $O_h$ phase at zero $\Omega$, and $C_{3v}(\alpha)$, $C_{3v}(\beta)$ and $C_{\infty v}$ phase for $0<\Omega/E_\text{L}<0.101$, $0.101<\Omega/E_\text{L}<0.133$ and $\Omega/E_\text{L}>0.133$, respectively. The phases are labelled by symmetry in the dressed state basis.  (c) Schematic of a ``Majorana spinor helix" in $C_{3v}(\beta)$ phase of (b). The grey lines are the trajectory of each Majorana point on the sphere versus $\hat{y}$ translation.      \label{SOC} }
\end{figure}

{\it SO-coupled regime and Majorana spinor helix:} The single-particle dispersion exhibits multiple minima that are almost degenerate in the SO-coupled regime. This occurs for small $\delta$ and $\Omega$.  
For $\Omega=0$, Eq.~(\ref{HDy}) exhibits $2F+1$ degenerate minima with wavefunctions $e^{im2k_{\text{L}}y}\varphi_m$ where $F_z\varphi_m=m\varphi_m$.
For small $\Omega$, the $2F+1$ local minima remain, but their energies $\epsilon_{m}$ are no longer degenerate and  
their wavefunctions are modified to (in leading order of $\Omega$):
\begin{equation}
\psi_m=e^{im2k_L y}\left(-\kappa_{m-1}\varphi_{m-1}+\varphi_m-\kappa_{m}\varphi_{m+1}\right),
\end{equation}
where $\kappa_m=\Omega\langle m+1|\hat{\Lambda}|m\rangle$.  In the presence of interactions, the ground state will be a linear combination these modified minima $\Phi=\sum_{m}\xi_m\psi_m$.  Expressing $\Phi$ in the original basis $\varphi_m$, which are spin components measured in a Stern-Gerlach experiment, we have 
$\Phi=\sum_m\eta_m\varphi_m$, where
\begin{equation}
\eta_m=e^{im2k_Ly}(-e^{-i2k_Ly}\kappa_{m-1}\xi_{m-1}+\xi_m-e^{i2k_Ly}\kappa_{m}\xi_{m+1}). \label{varphi}
\end{equation}
The coefficients $\xi_m$ 
can be determined in a straightforward fashion by minimizing the energy
$\mathcal{E}=\sum_{m=-F}^{F}\epsilon_m |\xi_m|^2+\mathcal{E}_{\text{int}}(\{\xi_{m}\})$,  where $\mathcal{E}_{\text{int}}(\{\xi_{m}\})$ is the interaction energy 
of the form
\begin{align}
&\mathcal{E}_{\text{int}}=\int d^3{\bf r}\sum_{j=0}^{F}g_{2j}\sum_{m=-j}^{j}\Pi^*_{jm}\Pi_{jm},\nonumber\\
&\Pi_{jm}=\sum_{m^\prime}\langle 2J,m|J,m^\prime,J,m-m^\prime\rangle\eta_{m^\prime}\eta_{m-m^\prime}, \label{Eint}
\end{align}
and $g_{2j}$ are $j+1$ independent interaction parameters in different total spin channels \cite{Ho}, and $\eta_{m}$ is related to $\xi_{m}$ via Eq.~(\ref{varphi}). The ground state is sensitive to variation of $\Omega$ since Raman coupling effectively changes the interaction between dressed states; in Fig.~\ref{SOC}(b) we give an example of the change of symmetry of the ground state as a function of $\Omega$  \cite{dipole}.

Equation~(\ref{varphi}) shows that each spin component of the ground state $\Phi=\sum_m\eta_m\varphi_m$ is a periodic function in $y$ with wavevector $k_{L}$. Such variations can not be described by spin rotations. Rather, it corresponds to each of the $2F$ Majorana points tracing out  different loops as the atoms travel over a period $\pi/k_{L}$ along $\hat{y}$, as shown in Fig.~3(c). We name such a structure a ``Majorana spinor helix," as opposed to a reshuffling of different Majorana points after a period, which would be (in the analog of superfluid $^{3}$He) a Majorana ``soliton.'' 

{\em Concluding remarks:} The outstanding challenge for experimental research employing a synthetic gauge field is the reduction of heating. Our discussion show that this problem can be solved by using atoms with large orbital and spin angular momentum such as Dy, which also has the advantage of considerably increasing the strength of the synthetic gauge field. Moreover, SO-coupled high-spin bosons and fermions possess far richer classes of broken symmetry ground states than  spin-1/2 fermions and spin-0 and spin-1 bosons. The realization of a strong synthetic gauge field and SO-coupling in these systems of large-spin atoms will constitute a major step toward  exploring 
exotic correlated states of fundamental importance that are difficult, if not impossible, to realize in solids.

We thank S.~Gopalakrishnan for early discussions and acknowledge support from NSFC under Grant No. 11104158 (X.C.), No. 11174176 (H.Z.), 11004118 (H.Z.), Tsinghua University Initiative Scientific Research Program (X.C. and T.L.H. and H.Z.), the NSF (B.L.L.), AFOSR (B.L.L.),  and NKBRSFC under Grant No. 2011CB921500 (H.Z.).  T.L.H. acknowledges the support by DARPA under the Army Research Office Grant Nos. W911NF-07-1-0464, W911NF0710576,  and by Tsinghua University through the Thousand People Plan.


\begin{thebibliography}{99}

\bibitem{Ian_magnetic}
Y. J. Lin, R. L. Compton, K. JimŽnez-Garc'a, J. V. Porto, and I. B. Spielman, Nature {\bf 462}, 628 (2009).

\bibitem{Ian_Hall}
L. J. LeBlanc, K. Jimenez-Garcia, R. A. Williams, M. C. Beeler, A. R. Perry, W. D. Phillips, and I. B Spielman, Proc. Natl. Acad. Sci. USA {\bf 109}, 10811 (2012).

\bibitem{Bloch_lattice}
M. Aidelsburger, M. Atala, S. Nascimb\'ene, S. Trotzky, Y. A. Chen, and I. Bloch, Phys. Rev. Lett. {\bf 107}, 255301 (2011).

\bibitem{Sengstock_lattice}
J. Struck, C. \"Olschl\"ager, R. Le Targat, P. Soltan-Panahi, A. Eckardt, M. Lewenstein, P. Windpassinger, and K. Sengstock, Science {\bf 333}, 996 (2011).

\bibitem{Ian_SO}
Y.-J. Lin, K. Jim\'{e}nez-Garc\'{\i}a, and I. B. Spielman, Nature \textbf{471}, 83 (2011).

\bibitem{Ian_partialwave}
R. A. Williams, L. J. LeBlanc, K. Jimenez-Garcia, M. C. Beeler, A. R. Perry, W. D. Phillips, and I. B. Spielman, Science {\bf 335}, 314 (2012).

\bibitem{Jing_SO}
Z. Fu, P. Wang, S. Chai, L. Huang, and J. Zhang, Phys. Rev. A {\bf 84}, 043609 (2011).

\bibitem{Shuai}
J. Y.  Zhang, S. C. Ji, Z. Chen, L. Zhang, Z. D. Du, B. Yan, G. S. Pan, B. Zhao, Y. Deng, H. Zhai, S. Chen, and J. W. Pan, Phys. Rev. Lett. {\bf 109}, 115301 (2012).

\bibitem{Jing_fermion}
P. Wang, Z. Q. Yu, Z. Fu, J. Miao, L. Huang, S. Chai, H. Zhai, and J. Zhang, Phys. Rev. Lett. {\bf 109}, 095301 (2012).

\bibitem{MIT_fermion}
L. W. Cheuk, A. T. Sommer, Z. Hadzibabic, T. Yefsah, W. S. Bakr, and M. W. Zwierlein, Phys. Rev. Lett. {\bf 109}, 095302 (2012).

\bibitem{Washtington}
C. Qu, C. Hamner, M. Gong, C. Zhang, and P. Engels, arXiv: 1301.0658

\bibitem{Zhai}
C. Wang, C. Gao, C.-M. Jian, and H. Zhai, Phys. Rev. Lett. \textbf{105}, 160403 (2010).

\bibitem{Ho_Zhang}
T.-L. Ho and S. Zhang, Phys. Rev. Lett. {\bf 107}, 150403 (2011).

\bibitem{Stringari}
Y. Li, L. P. Pitaevskii, and S. Stringari, Phys. Rev. Lett., \textbf{108}, 225301 (2012).

\bibitem{Stringari2}
Y. Li, G. L. Martone, and S. Stringari, Europhys. Lett., \textbf{99}, 56008  (2012).

\bibitem{Ian_PRA}
I. Spielman, Phys. Rev. A {\bf 79}, 063613 (2009).

\bibitem{Dy}
M. Lu, N. Q. Burdick, S.-H. Youn, and B. L. Lev, Phys. Rev. Lett. {\bf 107}, 190401 (2011).

\bibitem{Dy_fermion}
M. Lu, N. Q. Burdick, and B. L. Lev, Phys. Rev. Lett {\bf 108}, 215301 (2012).

\bibitem{Er}
K. Aikawa, A. Frisch, M. Mark, S. Baier, A. Rietzler, R. Grimm, and F. Ferlaino, Phys. Rev. Lett. {\bf 108}, 210401 (2012).

\bibitem{Majorana}
E. Majorana, Nuovo Cimento {\bf 9}, 43 (1932).

\bibitem{non-abelian}
R. Barnett, A. Turner, and E. Demler, Phys. Rev. Lett. {\bf 97}, 180412 (2006).

\bibitem{Schwinger}
R. Barnett, D. Poldolsky, and G. Refael, Phys. Rev. B {\bf 80}, 024420 (2009).

\bibitem{highspins}
A. Lamacraft, Phys. Rev. B {\bf 81}, 184526 (2010).

\bibitem{highspins2}
Y. Kawaguchi and M. Ueda, Phys. Rev. A {\bf 84}, 053616 (2011).



\bibitem{Biao}
B. Lian, T. L. Ho, and H. Zhai, Phys. Rev. A {\bf 85}, 051606(R) (2012).


\bibitem{ZhangHelix06}
B. A. Bernevig, J. Orenstein, and S.-C. Zhang, Phys. Rev. Lett. {\bf 97}, 236601 (2006).

 \bibitem{ZhangHelix09}
For instance, spin-$1/2$ spin helix has been observed in semiconductor quantum well, see J. D. Koralek, C. P. Weber, J. Orenstein, B. A. Bernevig, S.-C. Zhang, S. Mack and D. D. Awschalom, Nature {\bf 458}, 610 (2009) and reference therein.


\bibitem{Geremia}
J. M. Geremia, J. K. Stockton, and H. Mabuchi,  Phys. Rev. A {\bf 73}, 042112 (2006). 

\bibitem{deg}
This assumption can be satisfied at even small detunings from certain transitions in atoms like bosonic Dy. For instance, for bosonic $^{164}$Dy, the nuclear spin is zero and there is no quadratic Zeeman energy. Moreover, the 741-nm transition's excited state has nearly same $g$-factor as the ground state~\cite{LevSpect}. Thus, $\Delta_{ff^\prime}$ becomes independent of Zeeman shifts.

\bibitem{Deutsch}
I. H. Deutsch and P. S. Jessen, Phys. Rev. A {\bf 57}, 1972 (1998).

\bibitem{LevSpect}
M. Lu,  S.-H. Youn, and B. L. Lev, Phys. Rev. A {\bf 83}, 012510 (2011).


\bibitem{Cr}
The  tensor term is also not present in the SO-coupling hamiltonian for the spin-3  ${}^{52}$Cr atom, as has been derived by Y. Deng, J. Cheng, H. Jing, C.-P. Sun, and S. Yi, Phys. Rev. Lett. {\bf 108}, 125301 (2012). 

\bibitem{scalar}
The projected Hamiltonian also possesses an anti-trapping potential~\cite{Ian_PRA} and will have to be overcome with additional confinement from a far-off-resonance optical dipole trap laser. 



\bibitem{Ho}
T. -L. Ho, Phys. Rev. Lett. {\bf 81}, 742 (1998).

\bibitem{dipole} The effect of dipole-dipole interactions on these phase portraits remains to be explored.



\end{thebibliography}
\end{document}